\documentclass[aps,showpacs,twocolumn,superscriptaddress]{revtex4}
\usepackage{graphicx}
\usepackage{dcolumn}
\begin{document}

\title{Room for an $S=+1$ pentaquark in $K^+$ - nucleus phenomenology}

\author{A.~Gal}
\affiliation{Racah Institute of Physics, The Hebrew University,
Jerusalem 91904, Israel\vspace*{1ex}}

\author{E.~Friedman}
\affiliation{Racah Institute of Physics, The Hebrew University,
Jerusalem 91904, Israel\vspace*{1ex}}

\begin{abstract}
\rule{0ex}{3ex}

Evidence for excitation of exotic $S=+1$ pentaquark degrees of freedom
is presented by studying optical-potential fits to $K^+$ - nucleus
total, reaction and elastic-differential cross section data at
$p_{\rm lab} \sim 500 - 700$ MeV/c. Estimates of the underlying
two-nucleon absorption $K^+ nN \rightarrow \Theta^+ N$ reaction
cross section are made and are used for discussing the anticipated 
cross section of the strangeness exchange reaction 
$K^+ N \rightarrow \pi \Theta^+$.

\end{abstract}

\maketitle

\section{Introduction}
\label{sec:int}

Conclusive direct evidence for the existence of an exotic $S$=1, $I$=0,
$Z$=1 pentaquark baryon, the $\Theta^+$(1540) \cite{PDG04}, is still
lacking. Dedicated experiments using photons, pions and kaons are expected
to collect in due course sufficiently high statistics in order to resolve
this issue. However, in Ref. \cite{GF05} we have noted that
the $\Theta^+$(1540) provides a new mode of reactivity to $K^+$ - nuclear
interactions with possibly large effects on $K^+$ - nuclear total and
reaction cross sections in the energy range above its threshold
in nuclei, $p_{\rm lab}^{\rm th} \sim 400$ MeV/c. Since the $K^+N$
interaction in this energy range appears to be fairly weak and featureless,
without visible evidence for $KN \rightarrow \Theta^+$ coupling to exotic 
$qqqq\bar s$ configurations, 
pentaquark degrees of freedom in {\it nuclei} could be more readily excited
on two-nucleon clusters: $KNN \rightarrow \Theta^+ N$. 
This is related to the virtual two-meson decay mode
$\Theta^+ \rightarrow N K \pi$ \cite{CLM04,TRP05}.
In our earlier work we demonstrated how pentaquark production, corresponding
to the underlying two-nucleon absorption mode
\begin{equation}
\label{equ:nuc}
K^+ nN \rightarrow \Theta^+ N~,
\end{equation}
could contribute to the total and reaction cross sections 
\cite{FGM97a,FGM97b} extracted from transmission experiments 
\cite{WAA94,FGW97} at the Brookhaven National Laboratory (BNL) 
Alternating Gradient Synchrotron (AGS) on 
$^6$Li, $^{12}$C, $^{28}$Si and $^{40}$Ca at four energies, for
$p_{\rm lab} = 488, 531, 656, 714$ MeV/c. Our considerations are based 
merely on the observation \cite{FGM97a,FGM97b} that these $K^+$ - nucleus 
cross sections exhibit excessive reactivity, some $10 - 20\%$ over the 
reactivity provided by the $KN$ interaction, even after allowance 
is made for conventional nuclear medium effects. The suggestion that 
$S=+1$ pentaquark degrees of freedom give rise to this excess reactivity 
does not require that the $\Theta^+$ pentaquark has particular spin-parity 
values nor that it is as narrow as argued (less than $\Gamma \sim 1$ MeV) 
by analyzing $K^+$ initiated production processes \cite{CTr04,Gib04,SHK04}.

In the present work we provide a more detailed account of the calculations
presented briefly in Ref. \cite{GF05} for total and reaction cross sections,
extending these calculations to include also $K^+$ elastic scattering data
at $p_{\rm lab}=715$ MeV/c on $^6$Li and $^{12}$C \cite{MBB96,CSP97}.
Having determined the strength of the $K^+$ absorption mode
Eq. (\ref{equ:nuc}), we then discuss its relationship to the cross section
level expected for the $K^+ p \rightarrow \pi^+ \Theta^+$ production reaction 
which is under active experimentation in KEK at present \cite{Imai05}.

\section{Methodology}
\label{sec:meth}

The starting form adopted in our calculations for the kaon-nucleus 
optical potential $V_{\rm opt}$ is the simplest possible $t\rho$ form: 
\begin{equation}
\label{equ:Vopt}
2 \varepsilon^{(A)}_{\rm red} V_{\rm opt}(r) = -4\pi F_A b_0 \rho(r)~,
\end{equation}
where $\varepsilon^{(A)}_{\rm red}$ is the center-of-mass (c.m.) 
reduced energy, 
\begin{equation} 
\label{equ:kin1} 
(\varepsilon^{(A)}_{\rm red})^{-1}=E_p^{-1}+E_A^{-1} 
\end{equation} 
in terms of the c.m. total energies for the projectile and target 
respectively, and 
\begin{equation}
\label{equ:kin2}
F_A = \frac {M_A \sqrt{s}}{M(E_A+E_p)}
\end{equation}
is a kinematical factor resulting from the transformation
of amplitudes between the $KN$ and the $K^+$ - nucleus c.m. systems,
with $M$ the free nucleon mass, $M_A$ the mass of the target nucleus
and $\sqrt s$ the total projectile-nucleon energy in their c.m. system.
The parameter $b_0$ in Eq. (\ref{equ:Vopt}) reduces in the impulse 
approximation to the (complex) isospin-averaged $KN$ scattering
amplitude in the forward direction. 
The optical potential $V_{\rm opt}$ is inserted into 
the Klein Gordon equation, of the form used in our previous calculations: 
\begin{equation}
\label{equ:KG}
\left[ \nabla^2  + k^2 - (2 \varepsilon^{(A)}_{\rm red}
(V_{\rm c} + V_{\rm opt}) - {V_{\rm c}}^2) \right] \psi = 0 
\end{equation} 
in units of $\hbar = c = 1$. 
Here $k$ is the wave number in the c.m. system, and $V_{\rm c}$ is 
the Coulomb potential due to the charge distribution of the nucleus. 
These forms of the potential and the equation take into account $1/A$ 
corrections, which is an important issue when handling as light a nucleus 
as $^6$Li.

The nuclear density $\rho(r)$ is an essential ingredient of the
optical potential $V_{\rm opt}$ in Eq. (\ref{equ:Vopt}).
The density distribution of the protons is usually considered known as
it is obtained from the nuclear charge distribution \cite{FBH95} by
unfolding the proton charge distribution. For $^6$Li and for $^{12}$C
the modified harmonic oscillator (MHO) form was used whereas for $^{28}$Si
and for $^{40}$Ca the two-parameter Fermi (2pF) form was used.
In all cases the parameters were obtained numerically, requiring that
folding in the finite-size proton charge distribution will generate
a good fit to the nuclear charge distribution. For these $N=Z$ nuclei
we assumed in our previous analysis \cite{GF05} that the neutron densities
are identical to the corresponding proton densities. This choice is marked 
(i) in Table \ref{tab:rhobar}. In the present work
we have adopted also a slightly different approach for the two heavier
targets. Using the 2pF densities we obtained the parameters for the
proton distributions by approximate analytical unfolding of the charge
distribution \cite{PPe03}. The neutron densities were assumed to have an
`average' shape \cite{TJL01,FGa05} with a root-mean-square (rms)
radius $r_n$ given by $r_n-r_p=-0.0162A^{1/3}$~fm (only for $N=Z$ nuclei). 
This choice is marked (ii) in Table \ref{tab:rhobar}. 
By using two slightly different sets of densities we could test sensitivities
of the derived potential parameters.

\begin{table}
\caption{values of ${\bar \rho}$ (in fm$^{-3}$) Eq.~(\ref{equ:ave}), 
for the two sets of point-nucleon densities defined at the end of 
Sec.~\ref{sec:meth}.} 
\label{tab:rhobar} 
\begin{tabular}{ccccc} 
\hline \hline 
density & $^6$Li & $^{12}$C & $^{28}$Si & $^{40}$Ca \\  \hline 
(i)     &~ 0.049 ~&~ 0.104 ~&~ 0.112 ~&~ 0.112~ \\ 
(ii)    &~ 0.049 ~&~ 0.104 ~&~ 0.105 ~&~ 0.107~ \\ 
\hline \hline 
\end{tabular} 
\end{table}

\section{Results}
\label{sec:res}

\subsection{$K^+$ - nucleus total and reaction cross sections}

\begin{figure}[t]
\centerline{\includegraphics[height=6.8cm]{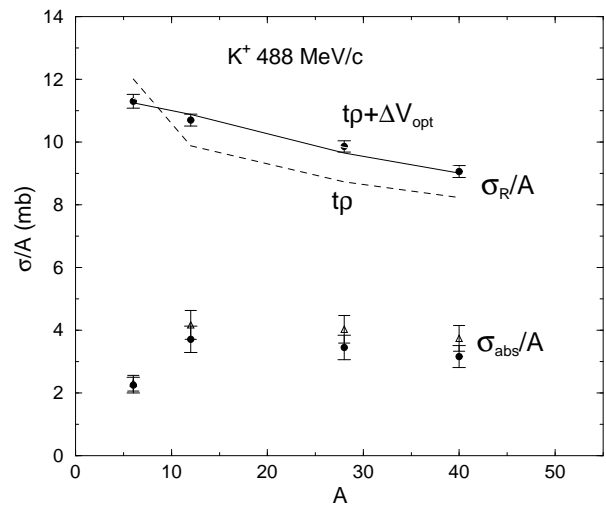}}
\caption{Data and calculations \protect\cite{GF05} for $K^+$ reaction cross
sections per nucleon ($\sigma_R/A$) at $p_{\rm lab}=488$ MeV/c are shown 
in the upper part. Calculated $K^+$ absorption cross sections per nucleon
($\sigma_{\rm abs}/A$) are shown in the lower part, see text.}
\label{kplusfig1}
\end{figure}

As reviewed recently in the Introduction of Ref. \cite{GF05}, the simple 
$t\rho$ form of $V_{\rm opt}$ in Eq. (\ref{equ:Vopt}) does not provide 
a satisfactory fit to the $K^+$ - nuclear integral cross section data at 
several hundreds of MeV. Indeed, using the methodology outlined above, 
it was shown by Friedman {\it et al.} \cite{FGM97a} that no 
{\it effective} value for $b_0$ could be found that fits satisfactorily 
the reaction and total cross sections derived from the BNL-AGS transmission 
measurements at $p_{\rm lab} = 488, 531, 656, 714$ MeV/c on $^6$Li, $^{12}$C, 
$^{28}$Si, $^{40}$Ca. This is demonstrated in the upper part of 
Fig. \ref{kplusfig1} for the reaction cross sections per nucleon 
$\sigma_R/A$ at 488 MeV/c, where the calculated cross sections using 
a best-fit $t\rho$ optical potential (dashed line) are compared with the 
experimental values listed in Ref. \cite{FGM97b}. The best-fit values of
Re~$b_0$ and Im~$b_0$ which specify this $t\rho$ potential are given in the
first row of Table \ref{tab:488}, where Im~$b_0$ represents $10-15\%$
increase with respect to the free-space value given in the line underneath. 
The $\chi ^2/N$ of this density-independent fit is very high. 
Its failure is due to the impossibility to reconcile the $^6$Li data 
(which for the total cross sections are consistent with the $K^+d$ 
`elementary' cross sections) with the data on the other, denser nuclei, 
as is clearly exhibited in Fig. \ref{kplusfig1} for the best-fit $t\rho$ 
dashed line. If $^6$Li is removed from the data base, then it becomes 
possible to fit reasonably well the data for the rest of the nuclei, 
but the rise in Im~$b_0$ with respect to its free-space value is then 
substantially higher than that for the $t\rho$ potential when $^6$Li is 
included. At the higher energies, $t\rho$ fits which exclude $^6$Li are 
less successful than at 488 MeV/c, while also requiring a substantial 
rise in Im~$b_0$, which means increased values of the in-medium $KN$ 
total cross sections with respect to the corresponding free-space values. 
This has been observed also in a $K^+$ - nucleus quasifree-scattering 
analysis \cite{Pet04}, for $K^+$ mesons incident on C, Ca, Pb at 
$p_{\rm lab}=705$ MeV/c \cite{KPS95}. 

\begin{table}
\caption{Fits to $K^+$ - nuclear integral cross sections 
\protect\cite{FGM97a,FGM97b} at $p_{\rm lab}=488$ MeV/c, 
using Eqs. (\ref{equ:KG}) - (\ref{equ:FGM97}).} 
\label{tab:488} 
\begin{tabular}{lllccc} 
\hline \hline 
$V_{\rm opt}$&Re~$b_0$ (fm)&Im~$b_0$ (fm)&$\beta$ (fm$^3$)&$\rho_{\rm th}$
(fm$^{-3}$) & $\chi ^2/N$  \\  \hline
$t\rho$ & $-$0.205(27) & 0.173(7) & -- &-- & 18.2 \\
$t_{\rm free}\rho$ & $-$0.178 & 0.153 & & & \\
Eq.~(\ref{equ:FGM97}) &$-$0.154(12)&0.160(2)&12.4(38)&0.088(6)& 0.06 \\
\hline \hline
\end{tabular}
\end{table}
 
An effective way of discriminating between $^6$Li and the denser nuclei
was established {\it empirically} in Refs. \cite{FGM97a,FGM97b} by requiring
that the imaginary (absorptive) part of the $K^+$ - nucleus optical potential
gets significantly enhanced whenever the average nuclear density
\begin{equation}
\label{equ:ave}
{\bar \rho}=\frac{1}{A}\int\rho^2d{\bf r}
\end{equation}
exceeds a threshold nuclear density $\rho_{\rm th}$. Specifically,
if Im~$V_{\rm opt}$ in Eq. (\ref{equ:Vopt}) is modified as follows,
\begin{equation}
\label{equ:FGM97}
{\rm Im} b_0~ \rho (r) \rightarrow {\rm Im} b_0~ \rho (r)~
[1 + \beta~({\bar \rho}-\rho_{\rm th})~\Theta({\bar \rho}-\rho_{\rm th})]~,
\end{equation}
then the long-standing problem of the reaction and total cross sections
derived from the BNL-AGS transmission measurements at
$p_{\rm lab} = 488, 531, 656, 714$ MeV/c on $^6$Li, $^{12}$C, $^{28}$Si,
$^{40}$Ca is resolved. This is demonstrated in Table \ref{tab:488},
showing two fits to the $N=8$ data points at 488 MeV/c, which is the 
closest momentum to the $\Theta^+$ resonance. The first fit in the table,
as explained above, uses only density-independent fitted values for the
complex parameter $b_0$ in Eq. (\ref{equ:Vopt}). The second fit in the
table introduces density dependence through Eq. (\ref{equ:FGM97}) and the
resulting improvement as judged by the value of $\chi ^2/N$ is spectacular.
When considering all 32 data points available at the 4 energies 
\cite{FGM97b}, using the same values for $\beta$ and
$\rho_{\rm th}$ independently of energy, then the $\chi ^2/N = 42.7$ for 
the best-fit $t\rho$ potential is reduced to 0.65 using this
empirical modification Eq. (\ref{equ:FGM97}).
The well-determined value of $\rho_{\rm th}$
is considerably higher than the average nuclear density $\bar \rho$ for
$^6$Li (about 0.05 fm$^{-3}$), but is lower than the $\bar \rho$ values
appropriate to the other, denser targets (about 0.1 fm$^{-3}$) as shown 
in Table \ref{tab:rhobar}. 
This spectacular fit clearly suggests that new absorptive degrees of freedom
open up above the threshold nuclear density of 0.09 fm$^{-3}$.
We have argued in Ref. \cite{GF05} that the $\Theta^+$ may 
provide for such a new degree of freedom via $K^+$ absorption on two
nucleons, $K^+nN \rightarrow \Theta^+N$, with threshold at
$p_{\rm lab}^{\rm th} \sim 400$ MeV/c.

Here we have incorporated $K^+nN \rightarrow \Theta^+N$ 
two-nucleon absorption into the impulse-approximation motivated 
$V_{\rm opt}(r)$, Eq. (\ref{equ:Vopt}), by adding a $\rho^2 (r)$ piece, 
as successfully practised in pionic atoms \cite{EEr66,BFG97} to account 
for $\pi^-$ absorption on two nucleons:
\begin{equation}
\label{equ:DD1}
b_0~ \rho (r) \rightarrow b_0~ \rho (r)~+~B~ \rho^2 (r)~,
\end{equation}
where the parameter $B$ represents the effect of $K^+$ nuclear absorption
into exotic $S=+1$ baryonic channels. Using this potential we have repeated
fits to all 32 data points for the reaction and total cross sections.
This resulted in a substantial improvement of the quality of the fit,
compared to the $t\rho$ potential. However, the fits at the higher
momenta are not as successful as the fit at 488 MeV/c, suggesting that
one needs a more effective way to distinguish between $^6$Li and the denser
nuclear targets. In fact, as demonstrated in Table \ref{tab:488} above,
the average nuclear density
${\bar \rho}$, Eq. (\ref{equ:ave}), provides for such discrimination
and is instrumental in achieving good agreement with experiment. 
We therefore replace Eq. (\ref{equ:DD1}) by the simplest ansatz
\begin{equation}
\label{equ:DD2}
b_0~ \rho (r) \rightarrow b_0~ \rho (r)~+~B~ {\bar \rho}~\rho (r)~.
\end{equation}
The added piece is a functional of the density which to lowest order
reduces to a $\rho^2$ form. Below we compare the two extensions of 
$V_{\rm opt}$ offered by Eqs. (\ref{equ:DD1}) and (\ref{equ:DD2}) and 
comment on the significance of the results obtained using the 
less founded form Eq. (\ref{equ:DD2}). 

\begin{table}
\caption{Fits to the eight $K^+$ - nuclear integral cross sections 
\protect\cite{FGM97b} at each of the four laboratory momenta $p_{\rm lab}$ 
(in MeV/c), using different potentials.}
\label{tab:FGa04}
\begin{tabular}{ccccccc}
\hline \hline
$p_{\rm lab}$&$V_{\rm opt}$&Re$b_0$(fm)&Im$b_0$(fm)&Re$B$(fm$^4$)&
Im$B$(fm$^4$)&$\chi^2/N$  \\  \hline
488&$t\rho$ &$-$0.203(26)&0.172(7)& & & 16.3 \\
   &$t_{\rm free}\rho$&$-$0.178&0.153 & & &  \\ 
   &Eq.(\ref{equ:DD1}) &$-$0.178&0.122(5)&0.52(20)&0.88(8) &1.18 \\
   &Eq.(\ref{equ:DD2}) &$-$0.178&0.129(4)&0.17(11)&0.62(6) &0.27 \\
 & & & & & & \\
531&$t\rho$ &$-$0.196(39)&0.202(9)& & & 56.3 \\
   &$t_{\rm free}\rho$&$-$0.172&0.170 & & &  \\ 
   &Eq.(\ref{equ:DD1}) &$-$0.172&0.155(14)&1.79(46)&0.72(27) &7.01 \\
   &Eq.(\ref{equ:DD2}) &$-$0.172&0.146(5) &0.46(21)&0.78(7) &3.94 \\
 & & & & & & \\
656&$t\rho$ &$-$0.220(50)&0.262(12)& & & 54.9 \\ 
   &$t_{\rm free}\rho$&$-$0.165&0.213 & & &  \\ 
   &Eq.(\ref{equ:DD1}) &$-$0.165&0.203(18)&1.66(80)&0.89(36) &7.24 \\
   &Eq.(\ref{equ:DD2}) &$-$0.165&0.204(5) &2.07(19)&0.77(7) &0.32 \\
 & & & & & & \\ 
714&$t\rho$ &$-$0.242(53)&0.285(15)& & & 67.7 \\ 
   &$t_{\rm free}\rho$&$-$0.161&0.228 & & &  \\ 
   &Eq.(\ref{equ:DD1}) &$-$0.161&0.218(24)&1.40(95)&1.10(48) &9.3 \\
   &Eq.(\ref{equ:DD2}) &$-$0.161&0.218(6)&1.51(43)&0.97(9)& 1.24 \\
\hline \hline
\end{tabular}
\end{table} 

Fits to the total and reaction cross section data \cite{FGM97b}, using 
Eqs. (\ref{equ:DD1}) and (\ref{equ:DD2}) with our set of slightly revised 
densities described above, are exhibited in Table \ref{tab:FGa04}. 
It is clear that the quality of fit improves dramatically with respect to 
the (also shown) $t\rho$ best fits upon allowing for $K^+$ absorption 
(parameter $B$). The superiority of the ${\bar \rho} \rho$ version
(marked as `Eq. (\ref{equ:DD2})') compared to the $\rho ^2$ version 
(marked as `Eq. (\ref{equ:DD1})') is also very clearly observed. 
The calculated reaction cross sections at 488 MeV/c,
using Eq. (\ref{equ:DD2}), are shown by the solid line marked
$t\rho + \Delta V_{\rm opt}$ in the upper part of Fig. \ref{kplusfig1},
where $\Delta V_{\rm opt}$ is the added piece of $V_{\rm opt}$ due to
a nonzero value of $B$. Clearly, it is a very good fit. 
Very recently, Tolos {\it et al.} \cite{TRP05} have demonstrated that
a similarly substantial improvement in the reproduction of reaction cross
sections could be achieved microscopically by coupling in degrees of
freedom of the $\Theta^+$(1540) pentaquark. 

We note that the splitting of Im~$V_{\rm opt}$ in Table \ref{tab:FGa04} 
into its two reactive components Im~$b_0$ and Im~$B$ appears well 
determined by the data at all energies, and perhaps is even model 
independent, particularly for the ${\bar \rho}\rho$ version 
Eq. (\ref{equ:DD2}) of the optical potential for which very accurate 
values of Im~$b_0$ are derived. These values of Im~$b_0$ are close to, 
but somewhat below the corresponding free-space values, in agreement 
with the conventional $t\rho \rightarrow g\rho$ medium effects considered 
in Ref. \cite{TRP05}.  
This is not the case for Re~$V_{\rm opt}$ where its two components 
are correlated strongly when Re~$b_0$ is varied too, largely cancelling 
each other into a resultant poorly determined Re~$V_{\rm opt}$. 
Therefore, in Table \ref{tab:FGa04} we show results only for Re~$b_0$ 
held fixed at its free-space value. 

Table \ref{tab:FGa04} suggests that 
the two-nucleon absorption coefficient Im~$B$ rises slowly with energy as 
appropriate to the increased phase space available to the underlying 
two-nucleon absorption process $K^+nN \rightarrow \Theta^+N$. 
Its values in this energy range are roughly independent of the form of 
$\Delta V_{\rm opt}$, the more conservative Eq. (\ref{equ:DD1}) 
or the more effective Eq. (\ref{equ:DD2}), used to derive these values 
from the data. This stability of the results for Im~$B$ is of special 
importance for the interpretation offered here. Regarding Re~$V_{\rm opt}$, 
and recalling that ${\bar \rho} \sim 0.1~ {\rm fm}^{-3}$ for the dense 
nuclear targets, it is clear that Re~$V_{\rm opt} \sim 0$ at the two higher 
momenta, illustrating the inadequacy of the $t\rho$ model which does not 
produce this trend. We note that our $K^+nN \rightarrow \Theta^+N$ absorption 
reaction is related to the mechanism proposed recently in Ref. \cite{CLM04}
as causing strong $\Theta^+$ - nuclear attraction, based on $K\pi$ two-meson
cloud contributions to the self energy of $\Theta^+$ in nuclear matter.
However, it would appear difficult to reconcile as strong
$\Theta^+$ - nuclear attraction as proposed there with the magnitude of
Re~$B$ derived in the present work.

\subsection{$K^+$ elastic scattering differential cross sections}

\begin{figure}[t]
            \centerline{\includegraphics[height=6.8cm]{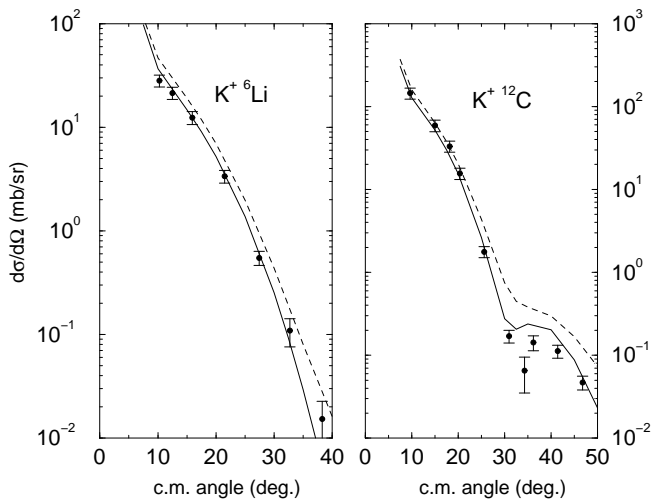}}
\caption{Comparison between measured differential cross sections for 
$K^+$ elastic scattering at $p_{\rm lab}=715$ MeV/c on $^6$Li and 
$^{12}$C \protect{\cite{CSP97}} and best-fit calculations using 
Eq.~(\ref{equ:DD1}) (dashed lines) and Eq.~(\ref{equ:DD2}) (solid lines).} 
\label{kplusfig2}
\end{figure}

In order to further test the picture that emerges from the analysis
of the integral cross sections for the $K^+$ - nucleus interaction,
we repeated the analysis including also differential cross sections for the
elastic scattering of $K^+$ by some of the  target nuclei.
Such data exist  for scattering of 715 MeV/c $K^+$ by $^6$Li and by
$^{12}$C \cite{MBB96,CSP97}. Similar data at 635 MeV/c were not included
because integral cross sections are not available at that energy.
Experience had shown that in situations where the real part
of the optical potential has a repulsive part, or at least is not
predominantly attractive, then fits to only angular distributions may
lead to values of potential parameters that result
in most unacceptable calculated values for reaction and total cross sections.
In other words, under such circumstances the integral cross sections serve
as powerful constraints on the potential parameters derived from fitting 
to differential cross sections \cite{MFJ89}.

Fits were made to the combined integral and differential cross sections
at 714 MeV/c consisting of the eight integral cross sections and the 17 
differential cross sections from Ref. \cite{CSP97}, using the potentials
of either Eq.~(\ref{equ:DD1}) or Eq.~(\ref{equ:DD2}). For the latter potential
and for the combined 25 data points we obtained $\chi ^2/N=5.3$, with
$\chi ^2/N=1.4$ for the integral cross sections and $\chi ^2/N=6.8$
for the differential cross sections. Using Eq.~(\ref{equ:DD1}) instead, 
we get a considerably inferior fit with $\chi ^2/N=24.6$. Potential 
parameters for Eq.~(\ref{equ:DD2}) are Re~$b_0$=$-$0.161 fm (fixed), 
Im~$b_0$=0.219$\pm$0.011 fm, Re~$B$=1.57$\pm$0.56 fm and 
Im~$B$=0.94$\pm$0.15~fm. These results are in perfect agreement with
the corresponding values in Table \ref{tab:FGa04}, obtained from fits
to integral cross sections only. Figure \ref{kplusfig2} shows comparisons
between calculations and experiment for the elastic scattering from
$^6$Li and $^{12}$C for the best-fit potentials for both Eq.~(\ref{equ:DD1})
(dashed) and Eq.~(\ref{equ:DD2}) (solid curves). The superiority of the 
${\bar \rho} \rho$ form is clear. 
Using the 14 differential cross section data of Ref. \cite{MBB96} instead of 
the 17 data points of Ref. \cite{CSP97}, together with the integral cross 
sections, leads to slightly lower values of $\chi ^2$ but the potential 
parameters are virtually the same as for the fits using the data of 
Ref. \cite{CSP97}.

\subsection{$K^+$ absorption cross sections}

By analogy to analyses of pionic atoms \cite{EEr66,BFG97} and low-energy 
pion-nuclear scattering reactions \cite{MFJ89,ASc86}, the additional 
piece $\Delta V_{\rm opt}$ due to the nonzero value of the absorption 
parameter $B$ is responsible for $K^+$ nuclear absorption into 
$\Theta^+$ - nuclear final states. 
One way to estimate the absorption cross section $\sigma_{\rm abs}^{(K^+)}$ 
is to use the distorted-wave Born approximation: 
\begin{equation}
\label{equ:abs1}
\sigma_{\rm abs}^{(K^+)} \sim  -~ {\frac{2}{\hbar v}}
\int {\rm Im} (\Delta V_{\rm opt}(r))~
|\Psi_{(\Delta V_{\rm opt}=0)}^{(+)}({\bf r})|^2~d{\bf r} ~~,
\end{equation}
where the distorted waves $\Psi_{(\Delta V_{\rm opt}=0)}^{(+)}$ are
calculated discarding $\Delta V_{\rm opt}$. Recall that for $B=0$,
replacing in the above integral $\Delta V_{\rm opt}(r)$ by $V_{\rm opt}(r)$
gives the total reaction cross section in the absence of the
$K^+nN \rightarrow \Theta^+N$ channel. However, the precise expression for
the total reaction cross section in the presence of this absorption mode
into $\Theta^+$ - nucleus final states requires the use of the fully
distorted waves $\Psi^{(+)}$, so that a different approximation for the
absorption cross section is given by
\begin{equation}
\label{equ:abs2}
\sigma_{\rm abs}^{(K^+)} \sim  -~ {\frac{2}{\hbar v}}
\int {\rm Im} (\Delta V_{\rm opt}(r))~
|\Psi^{(+)}({\bf r})|^2~d{\bf r} ~~.
\end{equation}
Calculated absorption cross sections {\it per target nucleon} at
$p_{\rm lab}=488$ MeV/c are shown in the lower part of Fig. \ref{kplusfig1}
for the fit using Eq. (\ref{equ:DD2}) for $V_{\rm opt}$ in Table
\ref{tab:FGa04}. The triangles are for expression (\ref{equ:abs1})
and the solid circles are for expression (\ref{equ:abs2}).
The error bars plotted are due to the uncertainty in the parameter Im~$B$.
It is seen that these calculated absorption cross sections,
for the relatively dense targets of C, Si and Ca, are proportional to the
mass number $A$, and the cross section per target nucleon due to
Im~$B \neq 0$ is estimated as close to 3.5 mb. Although the less 
successful Eq. (\ref{equ:DD1}) gives cross sections larger by $40\%$ 
at this particular incident momentum, this value
should be regarded an upper limit, since the best-fit density-dependent
potentials of Refs. \cite{FGM97a,FGM97b} yield values smaller than 3.5 mb
by a similar amount. 
The experience gained from studying $\pi$-nuclear absorption \cite{ASc86}
leads to the conclusion that $\sigma_{\rm abs}(K^+NN)$ is smaller than the
extrapolation of $\sigma_{\rm abs}^{(K^+)}/A$ in Fig. \ref{kplusfig1} to
$A=1$, and since the $KN$ interaction is weaker than the $\pi N$ interaction
one expects a reduction of roughly $50\%$, so that
$\sigma_{\rm abs}(K^+NN) \sim 1 - 2$ mb.

We note in Fig. \ref{kplusfig1} the considerably smaller absorption cross
section per nucleon calculated for $^6$Li which, considering its low density,
suggests a cross section of order fraction of millibarn for
$K^+ d \rightarrow \Theta^+ p$, well below the order 1 mb which as Gibbs has
argued recently \cite{Gib04} could indicate traces of $\Theta^+$ in $K^+ d$
total cross sections near $p_{\rm lab} \sim 440$ MeV/c. To be definite,
we suggested in Ref. \cite{GF05} the following range of values for this 
cross section: 
\begin{equation}
\label{equ:Kd}
\sigma(K^+ d \rightarrow \Theta^+ p)~ \sim ~ 0.1~-~0.5~ \rm{mb}~.
\end{equation} 
This provides a quantitative estimate for a possible missing-mass 
search for $\Theta^+$(1540) by observing the final proton, while 
the signal cross section need not exhibit a resonance 
behavior as function of the incoming $K^+$ momentum.

\section{Discussion} 
\label{sec:disc} 

We have provided a cross-section estimate, Eq. (\ref{equ:Kd}), for the 
{\it two-nucleon} production reaction $K^+ d \rightarrow \Theta^+ p$. 
It is worth emphasizing that this cross section is considerably
larger than what a {\it one-nucleon} production process
$KN \rightarrow \Theta^+$ would induce on a deuteron target.
Examples of one-step production processes in which the $\Theta^+$ 
is produced on one of the nucleons in a quasi on-shell kinematics are:
\begin{equation}
\label{equ:Theta1} 
K^+ ~p \rightarrow K^+ ~p ~, ~~~ K^+ ~n \rightarrow \Theta^+ ~,
\end{equation} 
\begin{equation} 
\label{equ:Theta2}
K^+ ~n \rightarrow K^0 ~p ~, ~~~ K^0 ~p \rightarrow \Theta^+ ~,
\end{equation} 
in which the $\Theta^+$ production is accompanied by initial scattering, or 
\begin{equation} 
\label{equ:Theta3} 
K^+ ~n \rightarrow \Theta^+ ~, ~~~ \Theta^+ ~p \rightarrow \Theta^+ ~p ~, 
\end{equation} 
in which it is followed by final scattering on the other `spectator' nucleon. 
The last process Eq. (\ref{equ:Theta3}) may be compared to a similar pion 
absorption process near the (3,3) resonance energy where the $\Delta$ 
is produced approximately on-shell, subsequently rescattering on the other 
nucleon, for example:
\begin{equation}
\label{equ:Delta} 
\pi^+ ~p \rightarrow \Delta^{++}~, ~~~ \Delta^{++}~n \rightarrow p~p ~,
\end{equation}
with a sizable cross section \cite{ASc86}
\begin{equation}
\label{equ:pid}
\sigma(\pi^+ d \rightarrow pp)~ \sim~ 12.5~ \rm{mb}~.
\end{equation}
Scaling by the ratio of coupling constants squared
$g^2_{KN\Theta}/g^2_{\pi N \Delta} \sim 2.5 \times 10^{-3}$, assuming
$J^{\pi}(\Theta^+) = {(\frac{1}{2})}^+$ and
$\Gamma (\Theta^+ \rightarrow K N) \sim 1$ MeV, we estimate a cross section
level of 0.03 mb for the one-step production process at the $\Theta^+$
resonance energy. [Assuming $J^{\pi}(\Theta^+) = {(\frac{1}{2})}^-$, the
one-step production cross section is lower by at least another order of
magnitude.] The one-step cross section affordable by the neutron Fermi
motion at $p_{\rm lab}=488$ MeV/c would be considerably smaller than this
estimate which holds at the very vicinity of the $\Theta^+$ mass for
$p_{\rm lab}=440$ MeV/c. In contrast, the two-nucleon reaction need not
involve the suppressed $KN\Theta$ coupling and its cross section which
we have estimated in Eq. (\ref{equ:Kd}) for $p_{\rm lab}=488$ MeV/c should
vary slowly with the kaon energy. The simplest mechanism for a two-nucleon
$K^+$ absorption process could be envisaged by letting an intermediate
off-shell pion correlate two target nucleons, viz.
\begin{equation} 
\label{equ:Kpi+} 
K^+ ~p \rightarrow \pi^+~\Theta^+~, ~~~\pi^+~n \rightarrow p ~,
\end{equation} 
or 
\begin{equation} 
\label{equ:Kpi0} 
K^+ ~n \rightarrow \pi^0~\Theta^+~, ~~~\pi^0~p \rightarrow p ~, 
\end{equation} 
where the threshold for the $K^+ N \rightarrow \pi\Theta^+$ reaction
occurs at $p_{\rm lab} \sim 760$ MeV/c in free space, and getting as low as 
$p_{\rm lab} \sim 550$ MeV/c in nuclear matter. This is a particular 
representation of the two-meson cloud contribution to the coupling of 
the $\Theta^+$ pentaquark in nuclei \cite{CLM04,TRP05}. Another process 
that does not depend directly on the suppressed $KN\Theta$ coupling 
involves the unknown $K^*N\Theta$ coupling constant: 
\begin{equation} 
\label{equ:K*+} 
K^+ ~p \rightarrow K^{*+}~p~, ~~~K^{*+}~~n \rightarrow p ~, 
\end{equation} 
or 
\begin{equation} 
\label{equ:K*0} 
K^+ ~n \rightarrow K^{*0}~p~, ~~~K^{*0}~~p \rightarrow p ~, 
\end{equation} 
with higher thresholds than for Eqs. (\ref{equ:Kpi+}) and (\ref{equ:Kpi0}). 
Estimates for the $K^+ p \rightarrow \pi^+\Theta^+$ reaction cross section 
suggest conservative values of order 0.1 mb \cite{OKL04,HHo05} which is 
of the scale needed to support a similar cross section level for 
$K^+ d \rightarrow \Theta^+ p$. 

\section{Summary and Conclusions}
\label{sec:conc}

There is a wide consensus that the impulse-approximation motivated $t\rho$ 
optical potential cannot reproduce the density dependence suggested by 
the $K^+$ - nuclear cross section data for incident momenta in the range 
$p_{\rm lab} \sim 450 - 800$ MeV/c. This was first realized by Siegel 
{\it et al.} \cite{SKG85} on the basis of old measurements of total cross 
sections and later on was reinforced \cite{CEr92,JEC95} using new 
transmission measurements data of total cross sections \cite{WAA94}. 
These investigations incorporated conventional medium effects such 
as off-shell dependence of the $KN$ $t$ matrix, Fermi averaging and the 
Pauli exclusion principle. The revised values of total cross sections, 
as well as the new reaction cross sections \cite{FGW97} which 
were subsequently extracted from these same transmission measurements, 
have led us together with Mare{\v s} \cite{FGM97a,FGM97b} to look for 
density dependence mechanisms that could resolve the striking discrepancy 
between experiment and theory. We have argued recently that the extra 
reactivity revealed by the $K^+$ - nucleus cross-section data is simply 
explained by adding a two-nucleon absorption channel 
$K^+ nN \rightarrow \Theta^+ N$ that couples in the $\Theta^+$(1540) 
pentaquark in a way which does not involve the apparently suppressed 
$KN\Theta^+$ coupling \cite{GF05}. The plausibility of this working 
hypothesis has been demonstrated very recently in Ref. \cite{TRP05} 
by evaluating the {\it unsuppressed} meson-cloud $K\pi N\Theta^+$ 
coupling which gives rise naturally to this two-nucleon absorption channel, 
with the same order of magnitude of $K^+$ absorption cross section 
as worked out by us \cite{GF05}. We wish to emphasize that this explanation 
does not require the assumed $S=+1$ pentaquark degrees of freedom 
to be materialized as a {\it narrow} $\Theta^+$(1540) $KN$ resonance, 
it only assumes that pentaquark degrees of freedom are spread over 
this energy range with sufficient spectral strength. 
If this is not the case, then the problem of excess reactivity in 
$K^+$-nuclear data remains unresolved, as demonstrated very recently 
by the new calculations of Ref. \cite{AVG05} (see in particular 
Figs. 9,10,and 12).  

In the present work, we have successfully reproduced the available
$K^+$ - nucleus integral (total as well as reaction) cross-section 
data on the four nuclear targets used in the energy range specified above 
\cite{WAA94,FGW97}, and also the elastic scattering angular distributions 
on $^6$Li and $^{12}$C at $p_{\rm lab}=715$ MeV/c \cite{MBB96,CSP97}, 
by adding to the $t\rho$ optical potential a density-dependent term which 
simulates absorption channels. The analysis of these data is consistent 
with an upper limit of about 3.5 mb on the $K^+$ absorption cross section 
per nucleon, for $\Theta^+$ production on the denser nuclei of $^{12}$C, 
$^{28}$Si, $^{40}$Ca, and indicates a sub-millibarn cross section for 
$\Theta^+$ production on deuterium. 
For a meaningful measurement of this $K^+ d \rightarrow \Theta^+ p$
two-body production reaction, an experimental accuracy of 0.1 mb in 
cross section measurements is required. It should provide a competitive 
production reaction to the $K^+  p \rightarrow \pi^+ \Theta^+$ 
two-body production reaction which is being measured at KEK \cite{Imai05}. 
For nuclear targets other than deuterium, given the magnitude
of the $K^+$ nuclear absorption cross sections as derived in the present
work, ($K^+,p$) experiments could prove useful. This reaction which has 
a `magic momentum' about $p_{\rm lab} \sim 600$ MeV/c, where the 
$\Theta^+$ is produced at rest, is particularly suited to study
bound or continuum states in {\it hyponuclei} \cite{Gol82}. It might prove
more useful than the large momentum transfer ($K^+,\pi^+$) reaction
proposed in this context \cite{NHO04}. 
In conclusion, precise low-energy $K^+d$ and $K^+$ - nuclear scattering 
and reaction data in the range $p_{\rm lab} \sim 300-800$ MeV/c, 
and particularly about 400 Mev/c, would be extremely useful to decide 
whether or not $S=+1$ pentaquark degrees of freedom are involved in the 
dynamics of $K^+$ - nuclear systems

\begin{acknowledgments}

This work was supported in part by the Israel Science Foundation grant 757/05.

\end{acknowledgments}


\begin{thebibliography}{99}

\bibitem{PDG04} G. Trilling in {\it Reviews of Particle Physics} edited by 
S. Eidelman {\it et al.} (Particle Data Group Collaboration), 
Phys. Lett. B {\bf 592}, 1 (2004).

\bibitem{GF05} A. Gal and E. Friedman, Phys. Rev. Lett. {\bf 94},
072301 (2005).

\bibitem{CLM04} D. Cabrera, Q.B. Li, V.K. Magas, E. Oset, and M.J. Vicente
Vacas, Phys. Lett. B {\bf 608}, 231 (2005).

\bibitem{TRP05} L. Tol{\'o}s, D. Cabrera, A. Ramos, and A. Polls, 
Phys. Lett. B {\bf 632}, 219 (2006).

\bibitem{FGM97a} E. Friedman, A. Gal, and J. Mare{\v s}, Phys. Lett. B
{\bf 396}, 21 (1997).

\bibitem{FGM97b} E. Friedman, A. Gal, and J. Mare{\v s}, Nucl. Phys. A
{\bf 625}, 272 (1997).

\bibitem{WAA94} R. Weiss, J. Aclander, J. Alster, M. Barakat, S. Bart, 
R.E. Chrien, R.A. Krauss, K. Johnston, I. Mardor, Y. Mardor, S. May Tal-beck, 
E. Piasetzky, P.H. Pile, R. Sawafta, H. Seyfarth, R.L. Stearns, R.J. Sutter, 
and A.I. Yavin, Phys. Rev. C {\bf 49}, 2569 (1994).

\bibitem{FGW97} E. Friedman, A. Gal, R. Weiss, J. Aclander, J. Alster, 
I. Mardor, Y. Mardor, S. May-Tal Beck, E. Piasetzky, A.I. Yavin, S. Bart, 
R.E. Chrien, P.H. Pile, R. Sawafta, R.J. Sutter, M. Barakat, K. Johnston, 
R.A. Krauss, H. Seyfarth, and R.L. Stearns, Phys. Rev. C {\bf 55}, 1304 
(1997).

\bibitem{CTr04} R.N. Cahn and G.H. Trilling, Phys. Rev. D {\bf 69}, 011501(R)
2004.

\bibitem{Gib04} W.R. Gibbs, Phys. Rev. C {\bf 70}, 045208 (2004).

\bibitem{SHK04} A. Sibirtsev, J. Haidenbauer, S. Krewald, 
and U.-G. Mei{\ss}ner, Phys. Lett. B {\bf 599}, 230 (2004); 
Eur. Phys. J. A {\bf 23}, 491 (2005).

\bibitem{MBB96} R.A. Michael, M.B. Barakat, S. Bart, R.E. Chrien, B.C. Clark, 
D.J. Ernst, S. Hama, K.H. Hicks, W. Hinton, E.V. Hungerford, M.F. Jiang, 
T. Kishimoto, C.M. Kormanyos, L.J. Kurth, L. Lee, B. Mayes, R.J. Peterson, 
L. Pinsky, R. Sawafta, R. Sutter, L. Tang, and J.E. Wise, Phys. Lett. B 
{\bf 382}, 29 (1996).

\bibitem{CSP97} R.E. Chrien, R. Sawafta, R.J. Peterson, R.A. Michael, and
E.V. Hungerford, Nucl. Phys. A {\bf 625}, 251 (1997).

\bibitem{Imai05} Experiment E559 at KEK-PS, K. Imai, Abstract 1WD.00002, 
2005, 2nd Meeting of the Nuclear Physics Divisions of the APS and the JPS, 
Maui, Hawaii, September 2005 (http://www.aps.org/meet/HAW05). 

\bibitem{FBH95} G. Fricke, C. Bernhardt, K. Heilig, L.A. Schaller,
L. Schellenberg, E.B. Shera, and C.W. De Jager, At. Data Nucl. Data
Tables {\bf 60}, 177 (1995).

\bibitem{PPe03} J.D. Patterson and R.J. Peterson, Nucl. Phys. A {\bf 717},
235 (2003).

\bibitem{TJL01} A. Trzci\'{n}ska, J. Jastrz\c{e}bski, P. Lubi\'{n}ski,
F.J. Hartmann, R. Schmidt, T. von Egidy, and B. K{\l}os,
Phys. Rev. Lett. {\bf 87}, 082501 (2001).

\bibitem{FGa05} E. Friedman, A. Gal, and J. Mare{\v s}, Nucl. Phys. A 
{\bf 761}, 283 (2005). 

\bibitem{Pet04} R.J. Peterson, Nucl. Phys. A {\bf 740}, 119 (2004). 

\bibitem{KPS95} C.M. Kormanyos, R.J. Peterson, J.R. Shepard, J.E. Wise, 
S. Bart, R.E. Chrien, L. Lee, B.L. Clausen, J. Piekarewicz, M.B. Barakat, 
E.V. Hungerford, R.A. Michael, K.H. Hicks, and T. Kishimoto, Phys. Rev. C 
{\bf 51}, 669 (1995). 

\bibitem{EEr66} M. Ericson and T.E.O. Ericson, Ann. Phys. (NY) {\bf 36}, 
323 (1966). 

\bibitem{BFG97} C.J. Batty, E. Friedman, and A. Gal, Phys. Rep. {\bf 287}, 
385 (1997). 

\bibitem{MFJ89} O. Meirav, E. Friedman, R.R. Johnson, R. Olszewski, and
P. Weber, Phys. Rev. C {\bf 40}, 843 (1989).

\bibitem{ASc86} D. Ashery and J.P. Schiffer, Annu. Rev. Nucl. Part. Sci.
{\bf 36}, 207 (1986).

\bibitem{OKL04} Y. Oh, H. Kim, and S.H. Lee, Phys. Rev. D {\bf 69}, 074016 
(2004). 

\bibitem{HHo05} T. Hyodo and A. Hosaka, Phys. Rev. C {\bf 72}, 055202 (2005). 

\bibitem{SKG85} P.B. Siegel, W.B. Kaufmann, and W.R. Gibbs, Phys. Rev. C
{\bf 31}, 2184 (1985). 

\bibitem{CEr92} C.M. Chen and D.J. Ernst, Phys. Rev. C {\bf 45}, 2019 (1992). 

\bibitem{JEC95} M.F. Jiang, D.J. Ernst, and C.M. Chen, Phys. Rev. C {\bf 51}, 
857 (1995). 

\bibitem{AVG05} H.F. Arellano and H.V. von Geramb, Phys. Rev. C {\bf 72}, 
025203 (2005). 

\bibitem{Gol82} A.S. Goldhaber, {\it Proceedings of the Second LAMPF II
Workshop}, edited by H.A. Thiessen, T.S. Bhatia, R.D. Carlini, and N. Hintz,
LA-9572-C, Vol. I (1982) 171.

\bibitem{NHO04} H. Nagahiro, S. Hirenzaki, E. Oset, and M.J. Vicente
Vacas, Phys. Lett. B {\bf 620}, 125 (2005).

\end{thebibliography}
\end{document}